\documentclass[11pt]{article} 

\usepackage[utf8]{inputenc} 

\usepackage{geometry} 
\usepackage{graphicx} 

\usepackage{booktabs} 
\usepackage{array} 
\usepackage{paralist} 
\usepackage{verbatim} 
\usepackage{subfigure} 
\usepackage{caption}
\usepackage{amsmath}
\usepackage{amssymb}
\usepackage{footnote}
\usepackage{float}
\usepackage{stackengine}
\usepackage{graphicx, array, blindtext}
\usepackage[toc,page]{appendix}
\usepackage{cite}

\usepackage{fancyhdr} 
\usepackage{sectsty}
\allsectionsfont{\sffamily\mdseries\upshape} 

\usepackage[nottoc,notlof,notlot]{tocbibind} 
\usepackage[titles,subfigure]{tocloft} 

\usepackage{authblk}

\title{Floquet analysis of self-resonance \\ in single-field models of inflation}
\author{Krzysztof Turzyński\thanks{krzysztof.turzynski@fuw.edu.pl}~ and Michał Wieczorek\thanks{michal.wieczorek@fuw.edu.pl}}
\affil{\small Institute of Theoretical Physics, Faculty of Physics, University of Warsaw, \\ Pasteura 5, 02-093 Warsaw, Poland}
\date{} 
\begin{document}
\maketitle
We review 52 models of single-field inflation,  paying special attention to the possibility that self-resonance of the unstable inflaton perturbations leads to reheating. 
We compute  Floquet exponents for the models that are consistent with current cosmological data. 
We find six models that exhibit a strong instability, but, barring the already known example of T-model of $\alpha$-attractors, only in one of them -- KKLT inflation -- the equation of state efficiently approaches that of radiation. 

\section{Introduction}

The concept of cosmological inflation~\cite{Starobinsky:1980te, Sato:1980yn, Guth:1980zm, Linde:1981mu, Albrecht:1982wi, Linde:1983gd}
has become a vital part of the standard cosmological model,
in particular, thanks to providing a mechanism for generating of primordial density perturbations~\cite{Starobinsky:1979ty, Mukhanov:1981xt, Hawking:1982cz,  Starobinsky:1982ee, Guth:1982ec, Bardeen:1983qw}.
However, after almost four decades of theoretical pursuit inflation remains a very general theory with no direct link to the Standard Model of particle physics 
(with a notable exception of so-called Higgs inflation \cite{HI1,HI2,HI3,HI4}). 
In particular, there is no universally accepted mechanism of reheating, {\it i.e.} the transition between inflationary era and radiation domination era.
Various plausible scenarios for reheating \cite{BT,KLS}, well embedded in the framework of quantum field theory have been thoroughly investigated (see, {\it e.g.}, \cite{Bassett:2005xm,Allahverdi:2010xz} for a review). 
One appealing possibility relies on mode amplification of quantum fluctuation and particle production that can occur when the homogeneous part of the inflaton field oscillates around 
the minimum of its potential \cite{PRH2,PRH3}. 
Quite recently, it has been understood that in some models such oscillations can excite fluctuations of the inflaton field to the extent that the Universe starts expanding as radiation dominated.
This mechanism, called self-resonance \cite{MA,AEF,AEFFH}, offers an economical and elegant exit from inflation in some inflationary models, greatly reducing theoretical uncertainty associated with reheating \cite{RE},
For a given model, one can numerically investigate self-resonance with lattice simulations, but this is both memory- and time-consuming and, so far, has been done only for a few inflationary models, see, {\it e.g.}, \cite{AL0,AL}. However, before undertaking full nonlinear lattice simulations one can approach the problem at linear order in perturbations 
and, using Floquet theory,  predict whether self-resonance can lead to efficient reheating (see, {\it e.g.}, \cite{FE} for a review).

In this letter, we utilize {\it Encyclopaedia Inflationaris}, a comprehensive review of single-field models of inflation \cite{EI}, 
to identify models which are both consistent with observational data (including the 2018 Planck data release \cite{PC}) and admit efficient self-resonance as the mechanism for reheating.
To this end, we employ Floquet analysis of inflaton perturbations.
The letter is organized as follows. In Section \ref{sec:two} we briefly present the self-resonance mechanism and briefly review Floquet theory. In Section \ref{sec:four}, we present the overview of single field inflationary models together with their prospective Floquet analysis. We draw our conclusions in Section \ref{sec:five}.

\section{Floquet analysis of self-resonance}
\label{sec:two}

\subsection{Inflationary perturbations}
\label{sec:twoa}

Throughout the paper we will consider the models of inflationary universe with one scalar field minimally coupled to gravity and with standard kinetic term, described by the action:
\begin{equation}\label{action}
S=\int \mathrm{d}^4x\sqrt{-g}\bigg[\frac{M_{P}^2}{2}R-\frac{1}{2}(\partial_\mu\phi)(\partial^\mu\phi)-V(\phi)\bigg],
\end{equation}
where 
\begin{equation}
\phi(t,\mathbf{x})\equiv\phi(t)+\delta\phi(t,\mathbf{x}).
\end{equation}
From now on, we will denote by $\phi$ the homogeneous part of the field. 
The metric $g_{\mu\nu}$ is the perturbed flat FLRW metric which in longitudinal gauge; in the absence of anisotropic stress it reads:
\begin{equation}\label{metric}
\mathrm{d}s^2=-(1+2\Psi)\,\mathrm{d}t^2+a^2(1-2\Psi)\,\mathrm{d}\mathbf{x}^2.
\end{equation}
Minimizing the action (\ref{action}), we obtain the following zeroth-order equations:
\begin{equation}\label{Friedmann}
H^2\equiv\left(\frac{\dot{a}}{a}\right)^2=\frac{1}{3M_{P}^2}\bigg[\frac{1}{2}\dot{\phi}^2+V\bigg],
 \end{equation} 
 \begin{equation}\label{first}
\ddot{\phi}+3H\dot{\phi}+V_{,\phi}=0.
\end{equation}
In the first order in perturbations, the relevant degree of freedom is the gauge invariant Mukhanov-Sasaki variable $Q\equiv\delta\phi+\frac{\dot{\phi}}{H}\Psi$, 
for which the equation of motion can be written as:
\begin{equation}\label{ruch1Q}
\ddot{Q}_k+3H\dot{Q}_k+\bigg(\frac{k^2}{a^2}+\mu_{\phi}^2\bigg)Q_k=0,
\end{equation}
where $Q(\mathbf{x},t)\equiv\int\frac{d^3\mathbf{k}}{(2\pi)^{3/2}}Q_k(t)e^{-i\mathbf{k}\mathbf{x}}$ and
\begin{equation}\label{muphi2}
\mu_{\phi}^2=\bigg(V_{,\phi\phi}-\frac{\dot{\phi}^4}{2M_{P}^4H^2} +3\frac{\dot{\phi}^2}{M_{P}^2}+2\frac{\dot{\phi}V_{,\phi}}{M_{P}^2H}\bigg)\, ,
\end{equation}
where the terms that involve $\dot{\phi}$ and inverse powers of the Planck mass arise due to the inclusion of the metric perturbations (see, {\it e.g.} \cite{FE}).
In certain models, solutions of (\ref{ruch1Q}) are unstable. Growing amplitudes of the Fourier modes of the inflaton perturbations indicate that energy is transferred from a homogeneous inflaton condensate to the inflaton fluctuations. This is the phenomenon called self-resonance \cite{MA,AEF,AEFFH}. In certain models, the kinetic and gradient energies of the inflaton fluctuations may eventually dominate the potential energy of the inflaton and the equation of state of the universe may approach that of radiation \cite{AL0,AL}.

\subsection{Floquet theorem}
In order to treat the possible growth of amplitude quantitatively, we will use Floquet analysis along the lines of Ref.~\cite{FE}.
To this end, we write the equation (\ref{ruch1Q}) as a set of two first-order equations:
\begin{equation}\label{Set}
\left(\begin{array}{c}
\dot{Q}_k \\
\dot{\Pi}_k 
\end{array}\right)=
\left(\begin{array}{ccc}
0 & 1 \\
-\bigg(\frac{k^2}{a^2}+\mu_{\phi}^2\bigg) & -3H
\end{array}\right)
\left(\begin{array}{c}
Q_k \\
\Pi_k 
\end{array}\right)
\end{equation}
After the end of inflation the field $\phi$ begins to oscillate around the minimum of the potential. These oscillations are almost periodic, as the changes of the period and the amplitude of the oscillations are slow compared to the time scale of the oscillations.
Moreover, the changes of $a(t)$ and $H(t)$ are also slow in this sense, as we show with numerical examples in Section~\ref{sec:four}. 
Therefore, the matrix on the r.h.s\ of eq.~(\ref{Set}) can be regarded as periodic.
Then, by Floquet theorem, the fundamental matrix $\mathcal{O}(t,t_0)$ of the solutions of the equation \eqref{Set} can be written as:
\begin{equation}\label{Floquet}
\mathcal{O}(t,t_0)=P(t,t_0)\exp{\Big[(t-t_0)\Lambda(t_0)\Big]}\,,
\end{equation} 
where $P(t,t_0)$ is a periodic matrix with the same period as the matrix in eq.~(\ref{Set}) and $P(t_0,t_0)$ is the identity matrix; $\Lambda(t_0)$ is a constant matrix with eigenvalues $\mu_k^{(i)2}$. A rule-of-thumb criterion for unstable growth of the amplitude of the inflaton perturbation is 
that the largest real part of the Floquet exponents
\begin{equation}
\label{eq:mumax}
\mu^{(\mathrm{max})}_k={}\mathrm{max}_i\{\mathrm{Re}(\mu_k^{(i)})\}
\end{equation}
is much larger than the Hubble scale, $\mu^{(\mathrm{max})}_k\gg H$.  
Our analysis deviates slightly from that of Ref.~\cite{FE}, as we include the Hubble friction, manifesting itself as $-3H$ entry in the matrix in (\ref{Set}) and calculate the average Floquet exponent for a few oscillations of $\phi$
between the turning points. A redefinition 
$\Pi = \tilde{\Pi}a^{-3}$ 
would remove the Hubble friction from the resulting equation of motion, which suggest that the sum of the two Floquet exponents is shifted by $-3H$.
The relation between the Floquet exponent calculated with and without the Hubble friction ($\mu_k$ and $\tilde{\mu}_k$, respectively) is model-dependent, as can be easily seen from the following two simple examples.
Let us assume for a while that the bottom left entry in the matrix on the right-hand side of (\ref{Set}) is just a negative constant $-\nu^2$. Then we have $\mathrm{Re}(\tilde{\mu}_1)=\mathrm{Re}(\tilde{\mu}_2)=0$; for $\nu^2=0$, we obtain $\mu_1=0$ and $\mu_2=-3H$, 
while for $\nu^2\gg H^2=\mathrm{const}$ we have $\mathrm{Re}(\mu_1)=\mathrm{Re}(\mu_2)=-\frac{3}{2}H$. Our numerical calculations suggest that most often the former possibility is realized: as we shall see, in models
that do not exhibit an instability the real part of one of the Floquet exponents is close to zero.
%
We also note that as $\tilde{\mu}_k^{(1)}+\tilde{\mu}_k^{(2)}=0$, the sign of $\mu^{(\mathrm{max})}_k$ can be either negative, zero
or positive, the latter signalling a potential unstable growth of the amplitude.

\subsection{Basics of Floquet analysis of the inflaton perturbations}

Eq.~(\ref{Floquet}) suggests that it is the maximal value of the real part of the Floquet exponent $\mu^{(\mathrm{max})}_k$ 
that is responsible for the growth of the perturbations, which increase as $\mathrm{exp}\left(\mu^{(\mathrm{max})}_k(t-t_0)\right)$. 
However, in the inflationary context
one also needs to take into account the width of the resonance band (the interval of wave numbers for which the Floquet exponents are positive) and the duration of the reheating era.
To quantitatively analyze this problem, we have to consider the evolution of perturbations during many periods of background oscillations, so the expansion of the Universe can no longer be neglected. Moreover, one has to take into account that the amplitude of the oscillations of the homogeneous inflaton field decreases due to Hubble friction. Therefore, considering the evolution of the perturbations through many background oscillations, one has to include the time dependence of Floquet exponents. It boils down to the following change of the function describing the growth of the amplitude:
\begin{equation}\label{Integral}
\exp\bigg[\mu^{(\mathrm{max})}_k\cdot(t-t_0)\bigg]\quad\rightarrow\quad \exp\bigg[\int_{t_0}^{t}\mathrm{d}t'\,\mu^{(\mathrm{max})}_k(t')\bigg],
\end{equation}
where $t_0$ now corresponds to the onset of the instability.

For an inflationary potential, which can be approximated as $V(\phi)\propto |\phi|^{2n}$ near its minimum, the time evolution of background oscillations amplitude can be described by the relation $\bar{\phi}\propto a^{-3/(n+1)}$. Simultaneously, in the expanding Universe the effective wave number decreases and satisfies $k_{\mathrm{eff}}=k/a$. Therefore, to track the time evolution of Floquet exponents, we can compute them for a range of values of the amplitude $\bar{\phi}$ and different wave numbers $k$. On the plane $(k,\bar{\phi})$, the end of inflation corresponds to a particular value of the amplitude of the homogeneous inflaton field ). As the Universe expands, the time evolution of the given mode corresponds to a particular path on the plane $(k,\bar{\phi})$ described by the relation $\bar{\phi}\propto k^{3/(n+1)}$. 
In Section \ref{sec:four}, we will present Floquet exponents calculated for various values of $k$ and $\bar{\phi}$,  the examples of the paths described above are marked with white lines and the end of the inflation
corresponds to red lines. 
The amplitude of a given mode will then grow according to (\ref{Integral}), in which Floquet exponents have to be evaluated at appropriate points along a path;
for significant growth, the relation $\mu^{(\mathrm{max}}_k(t))\gg H$ needs to be satisfied for approximately one Hubble time \cite{FE}. Empirically, we should usually require $\mu^{(\mathrm{max})}_k\gtrsim 10H$ (see, {\it e.g.}, \cite{AL}), which corresponds to a typical time during which a path crosses the instability patch.

The values of Floquet exponents are presented in units $M^2/M_{P}$. These units are natural for Floquet exponents for two reasons. First, the entire dependence of Floquet exponents on the scale $M\sim V^{1/4}$
is factored out.
Second, the Hubble rate is of order of $M^2/\sqrt{3}M_{P}$ at the end of inflation. Therefore, values of Floquet exponents describe naturally the rate of the exponential growth of the amplitude of a given mode: with \mbox{$\mathcal{M}_k\equiv\frac{\mu_k^{(\mathrm{max})}}{M^2/M_P}$}, the amplitude grows roughly by $\sim e^{\sqrt{3}\mathcal{M}_k}$ during one Hubble time. In these units the requirement for the strong resonance can be expressed as \mbox{$\mu_k^{(\mathrm{max})}\gtrsim 5M^2/M_P$}.

It is known that a strong growth of perturbations alone does not suffice to reheat the Universe, since the fragmented inflaton does not always acquire the equation of state characteristic for radiation. 
For example, very often oscillons are created
\cite{MA,AEF,AEFFH,AL0,AL,osc1,osc2,osc3,osc4,osc5,osc6,osc7,osc8,AS}, which constitute the dominant part of the energy in the Universe and yield the equation of state of non-relativistic matter. The persistence of oscillon-like solutions is typical for many inflationary potentials, which are quadratic near their minimum and flatten outside this region. As suggested in \cite{AL}, this fact seems to be quite general and is connected with the vanishing Floquet instability bands for $\bar{\phi}\rightarrow 0$ in such models.
Therefore, in our analysis we will focus on both the magnitude of Floquet exponents and the shape of the corresponding instability bands.

\section{Floquet analysis of single-field inflationary models}
\label{sec:four}

In the decades following the formulation of the inflationary scenario, numerous models of inflation have been proposed. 
It is a nearly Herculean task to describe them all comprehensively, because of both the huge number of authors who have worked on inflation and the shifting theoretical focus.
In our analysis of single-field inflationary models, we relied on a review by Martin, Ringeval and Vennin, aptly titled {\textit{Encyclopaedia Inflationaris} \cite{EI},} 
which provides a useful classification and parametrization of these models, as well as discussion about theoretical limits of their validity and references to original publications.
The computation of the Floquet exponents has been done for all the models that are in agreement with current cosmological data \cite{PC} for at least some values of the parameters.
Some inflationary models spurred theoretical interest in the past, 
but were later found to be inconsistent with data and are excluded from our analysis.
There are also models which were not included in Ref.~\cite{EI}, {\it e.g.}\ fibre inflation \cite{FI}, or theoretical ideas that have been proposed recently, {\it e.g.}\ $\alpha$-attractors \cite{Kallosh:2013yoa}, for which
we also perform the Floquet analysis. For completeness, let us also mention that the fate of perturbations in models inspired by string theory has been analyzed in \cite{Antusch:2017flz}, although none of
the examples presented therein points towards efficient reheating {\it via} self-resonance.



\subsection{Models for which Floquet analysis is impossible or irrelevant}

A number of inflationary models purports to describe only a limited part of the field range which is relevant for the generation of the perturbations that can be detected {\it via} CMB measurements.
Some of these models 
do not have a minimum with a vanishing value of the potential, so inflation cannot end properly.
There are 17 such models in Ref.~\cite{EI} and we excluded them from our analysis.
Further 4 models were excluded, because their potentials support slow-roll inflation for all field values, so
a graceful exit from these models of inflation must be provided by some additional mechanisms.

In order to obtain sufficient amount of inflation for GMSSMI, described by the potential
\begin{equation}
V(\phi)=M^4\bigg(\Big(\frac{\phi}{\phi_0}\Big)^2-\frac{2}{3}\alpha\Big(\frac{\phi}{\phi_0}\Big)^6+\frac{1}{5}\alpha\Big(\frac{\phi}{\phi_0}\Big)^{10}\bigg) \,,
\end{equation}
we need to choose a very fine-tuned value of the parameter $\alpha$, {\it i.e.}, it has to satisfy $0<1-\alpha\leq10^{-20}$. There are no strong theoretical arguments behind this fine-tuning, see, {\it e.g.}, \cite{GMSSMI1,GMSSMI2}. Moreover, even in case of the proper value of parameter $\alpha$, the predictions of this model are on the boundary of $2\sigma$ confidence level. With these two arguments against this model, we decided to forgo its Floquet analysis.




\subsection{Models with no large positive Floquet exponents}
\label{sec:small}

For all models listed in this Section, we perform the Floquet analysis described in Section~\ref{sec:two} and calculate the largest real part of the Floquet exponents (\ref{eq:mumax}) to
find that it is much smaller that the Hubble scale, so there is no unstable growth of the amplitude of inflaton perturbations in these models.

\begin{table}
\begin{center}
{\footnotesize
\begin{tabular}{|l|l|r|}
\hline
name and acronym & potential & \parbox{1.5cm}{maximal Floquet exponent} \\
\hline
Higgs inflation & $V(\phi)=M^4\bigg(1-e^{-\sqrt{2/3}\phi/M_{P}}\bigg)^2$ & $-5.2\times10^{-6}$\\
\hline
\end{tabular}
}
\caption{\it Models with no free parameters and with no large positive Floquet exponents \label{tab:nn0}}
\end{center}
\end{table}

Higgs inflation is the only zero-parameter model in this group. 
It should be stressed that there is a variety of theoretical ideas grouped under this label. The potential displayed in Table~\ref{tab:nn0} is the Einstein-frame potential of
the Starobinsky model \cite{Starobinsky:1980te}. It is also the effective single-field Einstein-frame description of the dynamic of a Higgs or Higgs-like field with a quartic potential and non-minimally coupled to gravity \cite{HI1,HI2,HI3,HI4}. 
 It is also possible to combine the two approaches into Higgs-Starobinsky model, see {\it e.g.} \cite{He:2018gyf}. We include the results of our Floquet analysis of this model for completeness, as
 more detailed studies of reheating within non-minimal Higgs field, taking account of the multi-field nature of the underlying model, exist in the literature \cite{HIR1,HIR2,HIR3}.

\begin{table}
\begin{center}
{\footnotesize 
\begin{tabular}{|l|l|l|r|}
\hline
name and acronym & potential & \parbox{1.7cm}{values of parameters} & \parbox{1.5cm}{maximal Floquet exponent} \\
\hline
\parbox{3cm}{$R+R^{2p}$ inflation (RpI)} & {\footnotesize $V(\phi)=M^4e^{-\sqrt{\frac{8}{3}} \frac{\phi}{M_{P}}}\Big|e^{\sqrt{\frac{2}{3}}\frac{\phi}{M_{P}}}-1\Big|^\frac{2p}{2p-1}$ } & $p=1.01$ & $-9.1\times10^{-2}$  \\
\hline
\parbox{3cm}{Large field inflation (LFI)} & {\footnotesize $V(\phi)=M^4\bigg(\frac{\phi}{M_{P}}\bigg)^p$ } & $p=2$ & $2.3\times10^{-4}$ \\
\hline
\parbox{3cm}{Mixed Large field inflation (MLFI)} & {\footnotesize $V(\phi)=M^4\bigg(\frac{\phi}{M_{P}}\bigg)^2\bigg(1+\alpha\frac{\phi^2}{M_{P}^2}\bigg)$ } & $\alpha=5\times10^{-4}$ & $3.0\times10^{-4}$  \\
\hline
\parbox{3cm}{Radiatively corrected massive inflation (RCMI)} & {\footnotesize $V(\phi)=M^4\frac{\phi^2}{M_{P}^2}\bigg(1-2\alpha\frac{\phi^2}{M_{P}^2}\mathrm{ln}\Big(\frac{\phi}{M_{P}}\Big)\bigg)$ } & $\alpha=10^{-4}$ & $3.4\times10^{-4}$  \\
\hline
\parbox{3cm}{Natural inflation (NI)} & {\footnotesize $V(\phi)=M^4\bigg(1+\cos\Big(\frac{\phi}{f}\Big)\bigg)$ } & $f=100M_P$ & $5.8\times10^{-6}$  \\
\hline
\parbox{3cm}{Kahler moduli inflation~I (KMII)} & {\footnotesize $V(\phi)=M^4\bigg(1-\alpha\frac{\phi}{M_{P}}e^{-\phi/M_{P}}\bigg)$ } & $\alpha=e$ & $-4.6\times10^{-3}$  \\
\hline
\parbox{3cm}{Double well potential inflation (DWI)} & {\footnotesize $V(\phi)=M^4\bigg(\Big(\frac{\phi}{\phi_0}\Big)^2-1\bigg)^2$ } & $\phi_0=25M_{P}$ & $1.1\times10^{-4}$  \\
\hline
\parbox{3cm}{Fibre inflation (FI)} & {\footnotesize $\begin{array}{ll} V(\phi) =& M^4\left( (3-R)-4(1-\frac{R}{6})e^{-\frac{\phi}{\sqrt{3}M_P}} +\right. \\ & \left.+(1+\frac{2R}{3})e^{-\frac{2\phi}{\sqrt{3}M_P}} +R e^{\frac{\phi}{\sqrt{3}M_P}}\right)\end{array}$ } & $R=10^{-6}$ & $-2.6\times10^{-6}$  \\
\hline
\end{tabular}
}
\caption{\it Models with one parameter and with no large positive Floquet exponents. Values of Floquet exponents are given in units of $M^2/M_P^2$. \label{tab:nn1}}
\end{center}
\end{table}

Predictions of the one-parameter models listed in Table~\ref{tab:nn1} agree with Planck data at $2\sigma$ confidence level and we can perform Floquet analysis for the inflaton field oscillating around the minimum of the potential. We used the values of the parameters that are favored by the data and justified by theoretical derivation of the models.
We also extended the list of models beyond Ref.~\cite{EI}, including fibre inflation put forth in Ref.~\cite{FI}.
For all the models in this group, the obtained Floquet exponents are negative or negligible.

\begin{table}
\begin{center}
{\footnotesize
\begin{tabular}{|l|l|l|r|}
\hline
name and acronym & potential & \parbox{1.7cm}{values of parameters} & \parbox{1.5cm}{maximal Floquet exponent}  \\
\hline
\parbox{3cm}{Supergravity brane inflation (SBI)} & $V(\phi)=M^4\bigg(1+\bigg(-\alpha+\beta\mathrm{ln}\Big(\frac{\phi}{M_{P}}\Big)\bigg)\Big(\frac{\phi}{M_{P}}\Big)^4\bigg)$ & \parbox{1.9cm}{\mbox{$\beta=10^{-4}$} $\alpha=\frac{\beta/4}{1-\ln(\beta/4)}$} & $6.7\times10^{-4}$ \\
\hline
\parbox{3cm}{Spontaneous symmetry breaking inflation (SSBI)} & $V(\phi)=M^4\bigg(1+\alpha\Big(\frac{\phi}{M_{P}}\Big)^2+\beta\Big(\frac{\phi}{M_{P}}\Big)^4\bigg)$ & \parbox{1.99cm}{\mbox{$\alpha=-2
\times 10^{-3}$} \mbox{$\beta=10^{-6}$} }& $1.0\times10^{-4}$ \\
\hline
\end{tabular}
}
\caption{\it Models with two parameters and with no large positive Floquet exponents. Values of Floquet exponents are given in units of $M^2/M_P^2$. \label{tab:nn2}}
\end{center}
\end{table}

Models with two parameters are listed in Table~\ref{tab:nn2}.
In SBI, the parameter $\beta\approx 10^{-4}$ gives the best agreement with Planck data, at the border of $2\sigma$ confidence level. To obtain the minimum of the potential at zero, we have to take \mbox{$4\alpha=\beta(1-\mathrm{ln}(\beta/4))$}, so there remains just one adjustable parameter. We performed Floquet analysis with these parameters values.
For the SSBI model, it is convenient
to divide the analysis into the following cases: \par $1)$ $\alpha>0$, $\beta>0$ \par$2)$ $\alpha<0$, $\beta<0$ \par$3)$ $\alpha>0$, $\beta<0$, $\phi^2_{\rm start}<-\alpha/2\beta$ \par$4)$ $\alpha>0$, $\beta<0$, $\phi^2_{\rm start}>-\alpha/2\beta$ \par$5)$ $\alpha<0$, $\beta>0$, $\phi^2_{\rm start}<-\alpha/2\beta$ \par$6)$ $\alpha<0$, $\beta>0$, $\phi^2_{\rm start}>-\alpha/2\beta$, \\
where $\phi_{\rm start}$ denotes the value of the inflaton field at the beginning of observable inflation.
We performed Floquet analysis for cases 3) and 5) and found no large positive Floquet exponents. Other cases either do not agree with Planck data at $2\sigma$ confidence level or 
the potential has no minimum at zero.

\begin{table}
\begin{center}
{\footnotesize
\begin{tabular}{|l|l|l|r|}
\hline
name and acronym & potential & \parbox{3.2cm}{values of parameters} & \parbox{1.5cm}{maximal Floquet exponent} \\
\hline
\parbox{3cm}{Logarithmic potential inflation (LPI)} & $V(\phi)=M^4\bigg(\frac{\phi}{\phi_0}\bigg)^p\bigg(\log\frac{\phi}{\phi_0}\bigg)^q$ & \parbox{3.2cm}{\mbox{$p=3,q=4,\phi_0=M_P$} \mbox{$p=2,q=2,\phi_0=M_P$}} & \parbox{1.5cm}{$1.9\times10^{-5}$ $3.6\times10^{-5}$}\\
\hline
\end{tabular}
}
\caption{\it Models with three parameters and with no large positive Floquet exponents. Values of Floquet exponents are given in units of $M^2/M_P^2$. \label{tab:nn3}}
\end{center}
\end{table}

There is just a single three-parameter model, LPI, in this category; we present it in Table~\ref{tab:nn3}. 
There are different regions of this potential, where inflation can hold. However all of them are either disfavored by Planck data or the obtained Floquet exponents are negligibly small to support self-resonance reheating in case of this model.

\subsection{Models with large positive Floquet exponents}
\label{sec:large}

\begin{table}
\begin{center}
{\footnotesize
\begin{tabular}{|l|l|l|r|}
\hline
name and acronym & potential & \parbox{3.2cm}{values of parameters} & \parbox{1.5cm}{maximal Floquet exponent} \\
\hline
\parbox{3.5cm}{Mutated hilltop inflation (MHI)} & $V(\phi)=M^4\bigg(1-\mathrm{sech}\Big(\frac{\phi}{\mu}\Big)\bigg)$ &
\parbox{3cm}{$\mu=0.02M_P$\\$\mu=0.04M_P$} & \parbox{1.5cm}{11 \\ 5.2 }
\\
\hline
\parbox{3.5cm}{Radion gauge inflation (RGI)} & $V(\phi)=M^4\frac{(\phi/M_{P})^2}{\alpha+(\phi/M_{P})^2}$ &
\parbox{3cm}{$\alpha=0.001$ \\ $\alpha=0.002$} & \parbox{1.5cm}{ 3.7 \\ 6.6 }
\\
\hline
\parbox{3.5cm}{Witten-O'Raifeartaigh inflation (WRI)} & $V(\phi)=M^4\mathrm{ln}^2\Big(\frac{\phi}{\phi_0}\Big)$ &
\parbox{3cm}{$\phi_0=0.01 M_{P}$ \\ $\phi_0=0.02 M_{P}$} & \parbox{1.5cm}{11 \\ 4.9}
\\
\hline
\end{tabular}
}
\caption{\it Models with one parameter and large positive Floquet exponents for given values of parameters. Values of Floquet exponents are given in units of $M^2/M_P^2$. \label{tab:ok1}}
\end{center}
\end{table}

\begin{figure}
\centering
\begin{tabular}{cc}
$\mu = 0.04$ & $\mu = 0.02$ \\
\includegraphics[width=0.48\textwidth, height=0.35\textwidth]{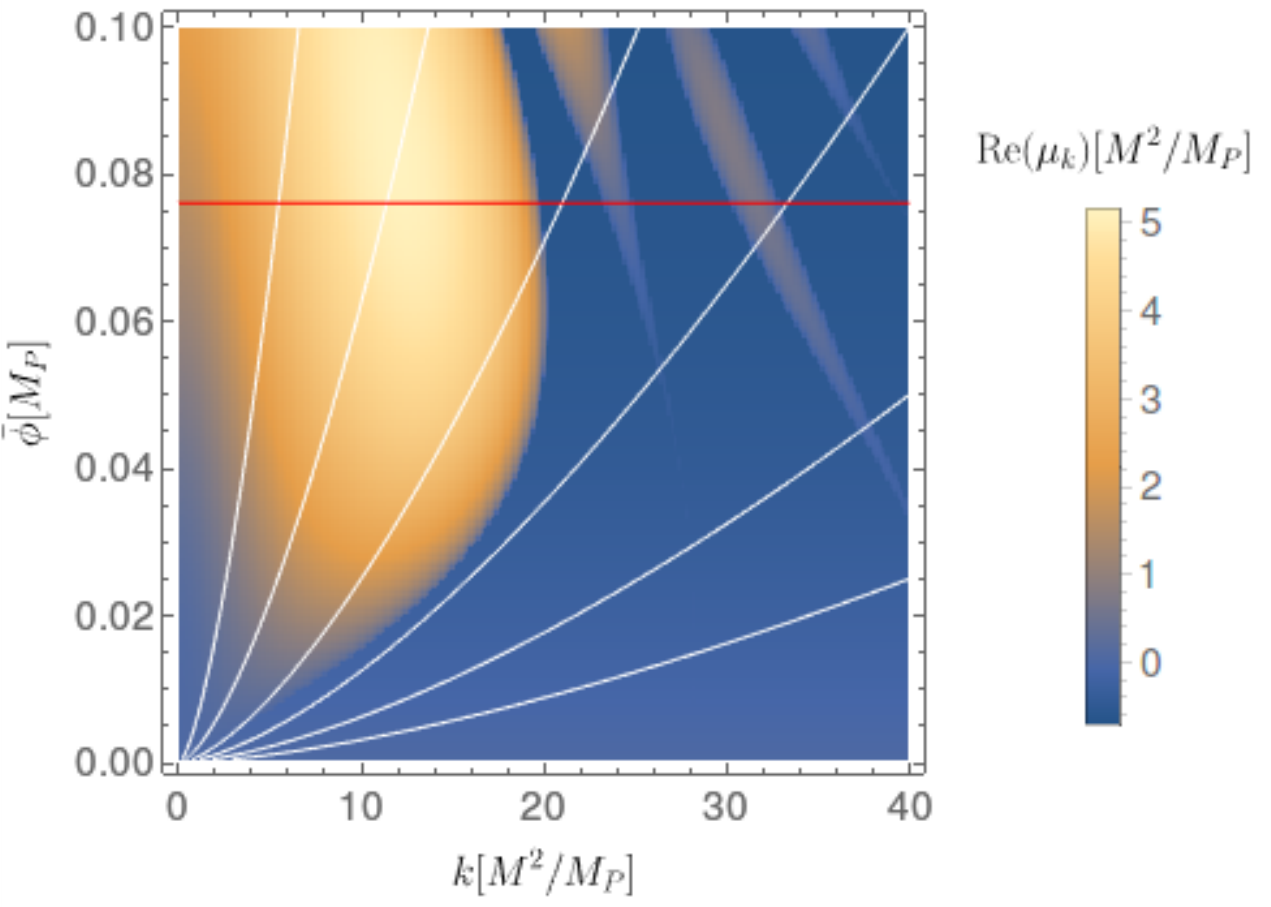} &
\includegraphics[width=0.48\textwidth, height=0.35\textwidth]{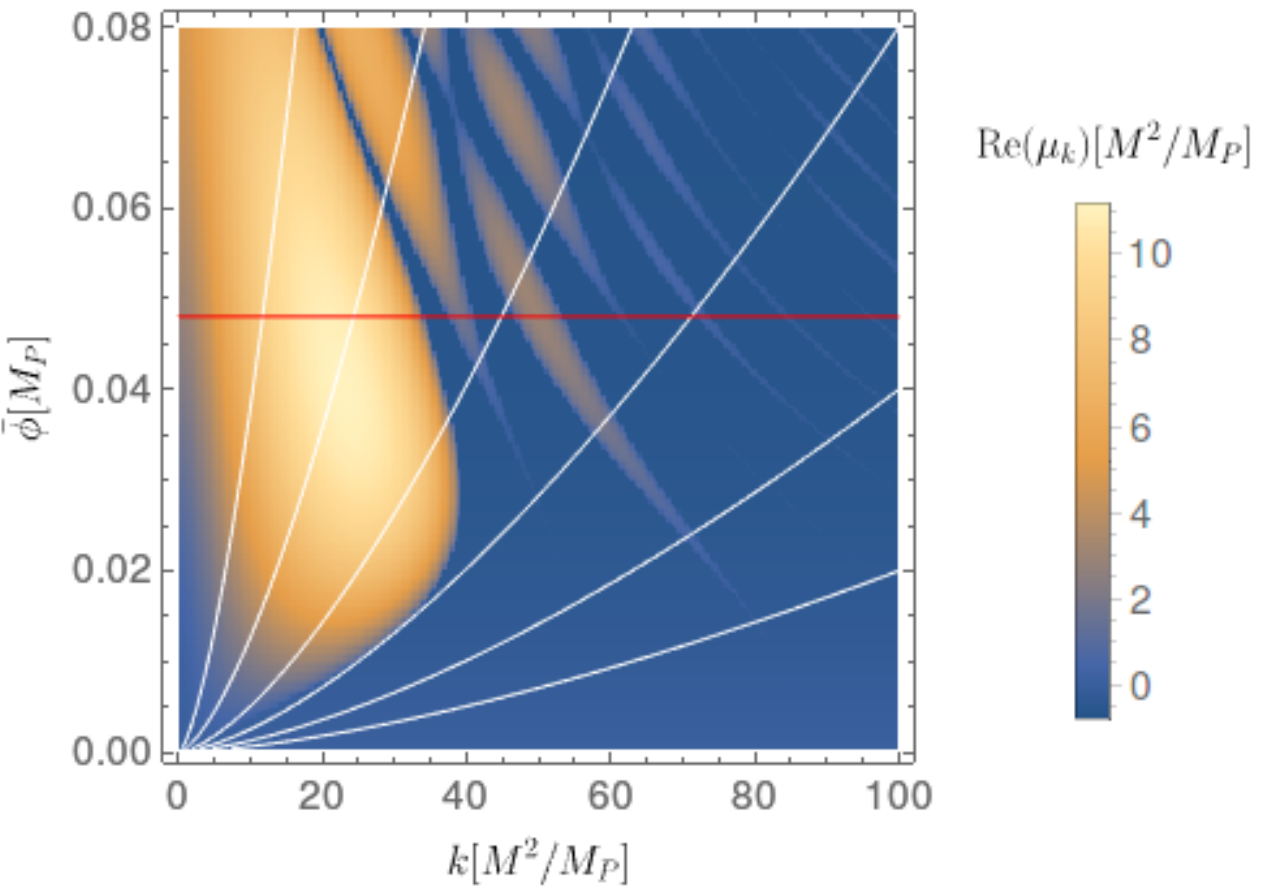}
\end{tabular}
\caption{\it Floquet exponents for the model MHI
for two different parameter choices consistent with Planck data.
  \label{fig:mhi}}
\end{figure}

\begin{figure}
\centering
\begin{tabular}{cc}
$\alpha = 0.004$ & $\alpha = 0.002$\\
\includegraphics[width=0.48\textwidth, height=0.35\textwidth]{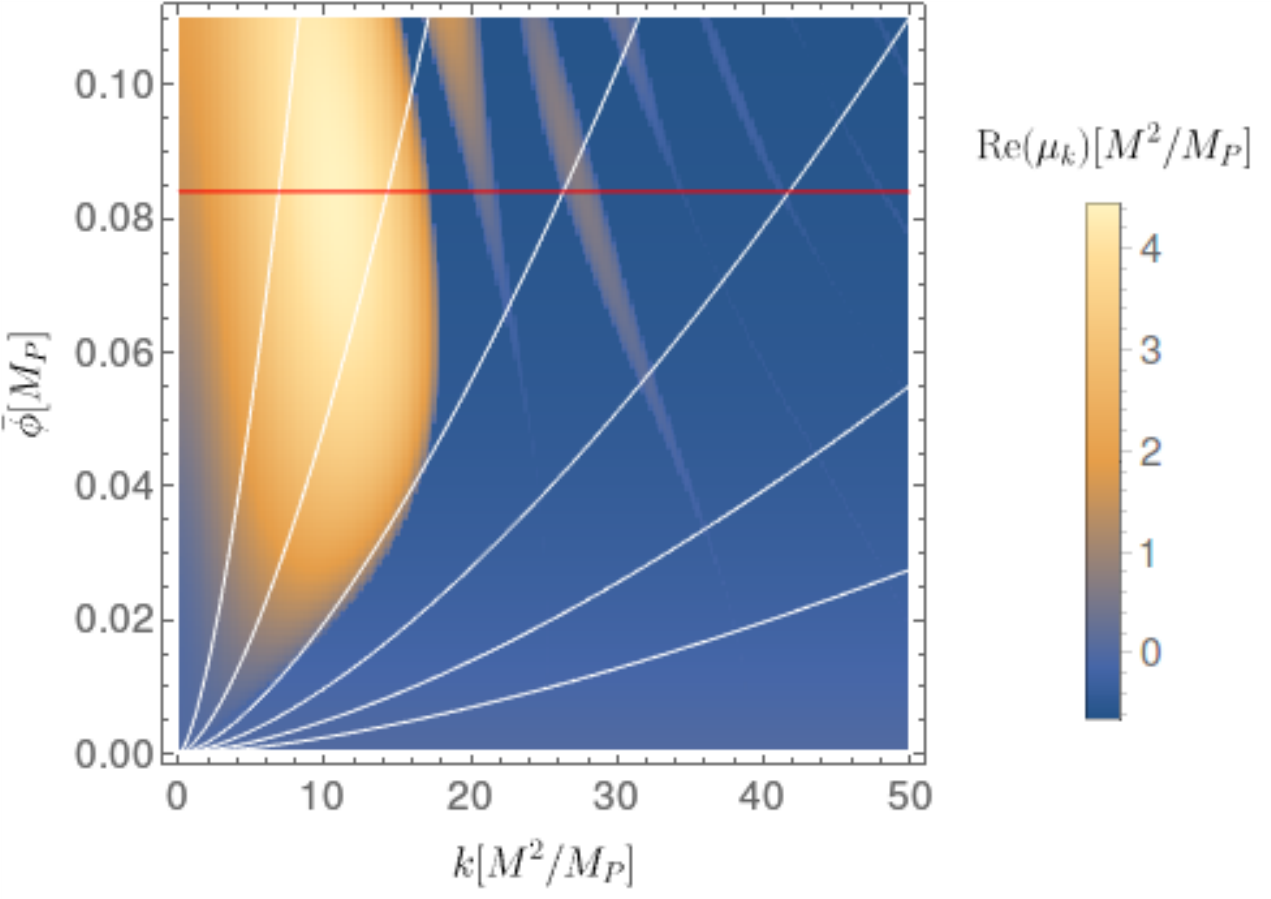} &
\includegraphics[width=0.48\textwidth, height=0.35\textwidth]{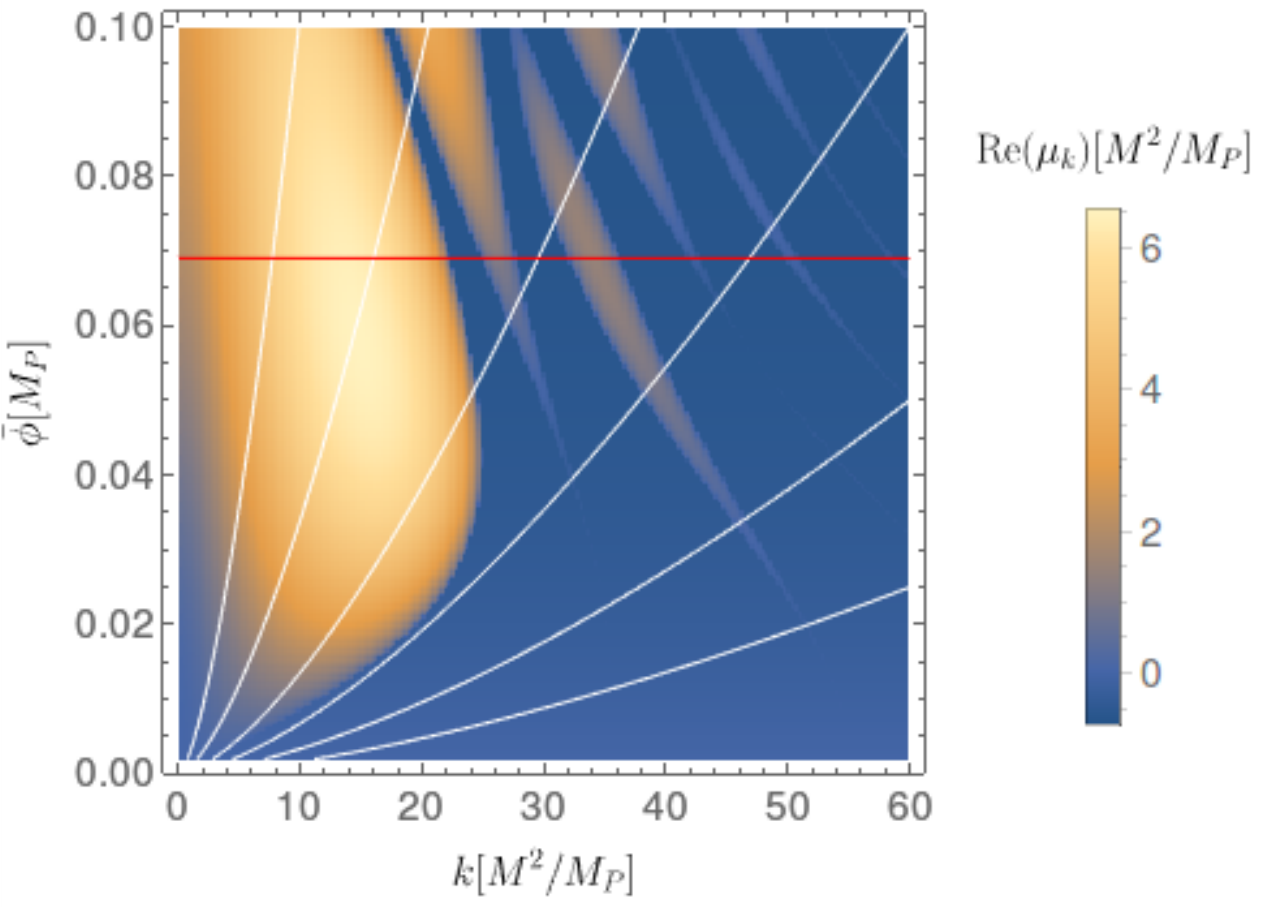}
\end{tabular}
\caption{\it Floquet exponents for the model RGI
for two different parameter choices consistent with Planck data.
  \label{fig:rgi}}
\end{figure}

\begin{figure}
\centering
\begin{tabular}{cc}
$\phi_0 = 0.01$ & $\phi_0 = 0.02$\\
\includegraphics[width=0.48\textwidth, height=0.35\textwidth]{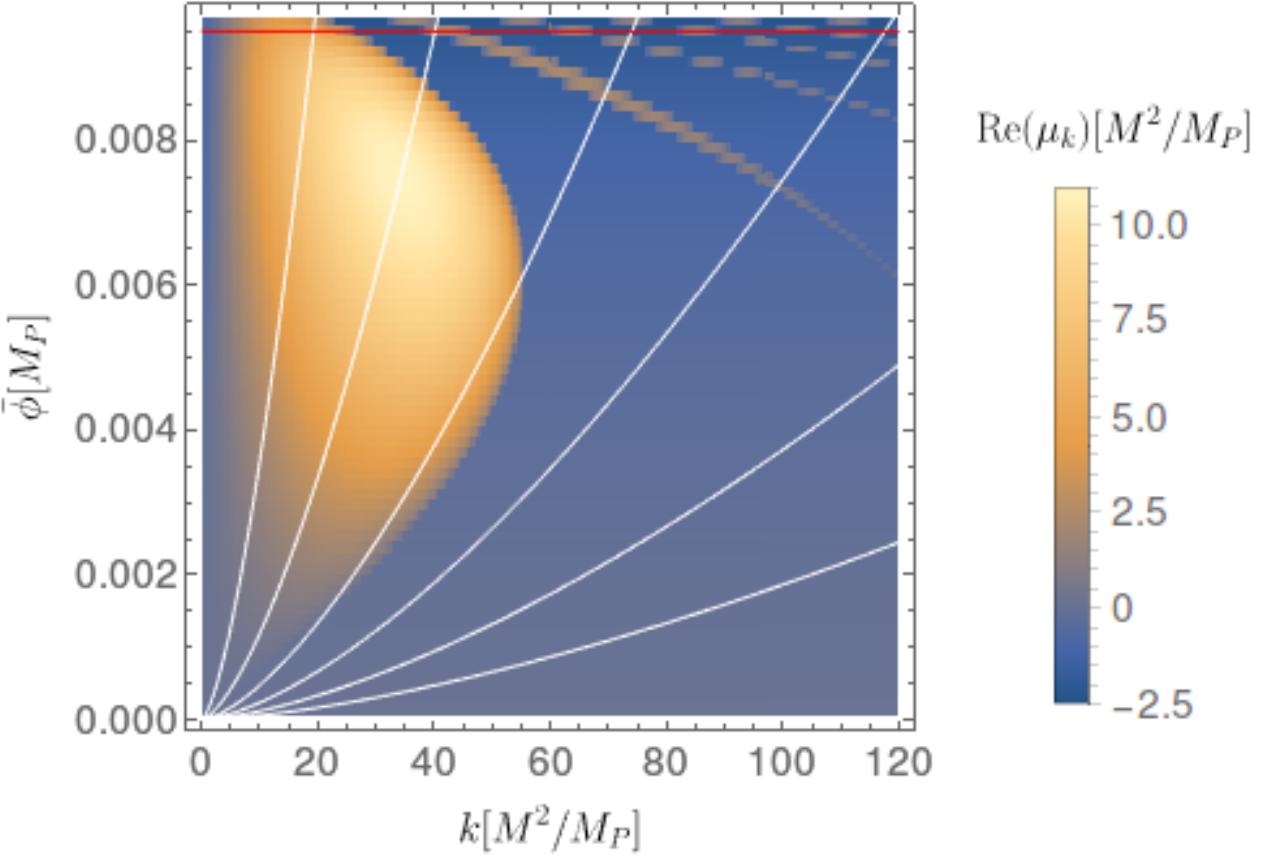} &
\includegraphics[width=0.48\textwidth, height=0.35\textwidth]{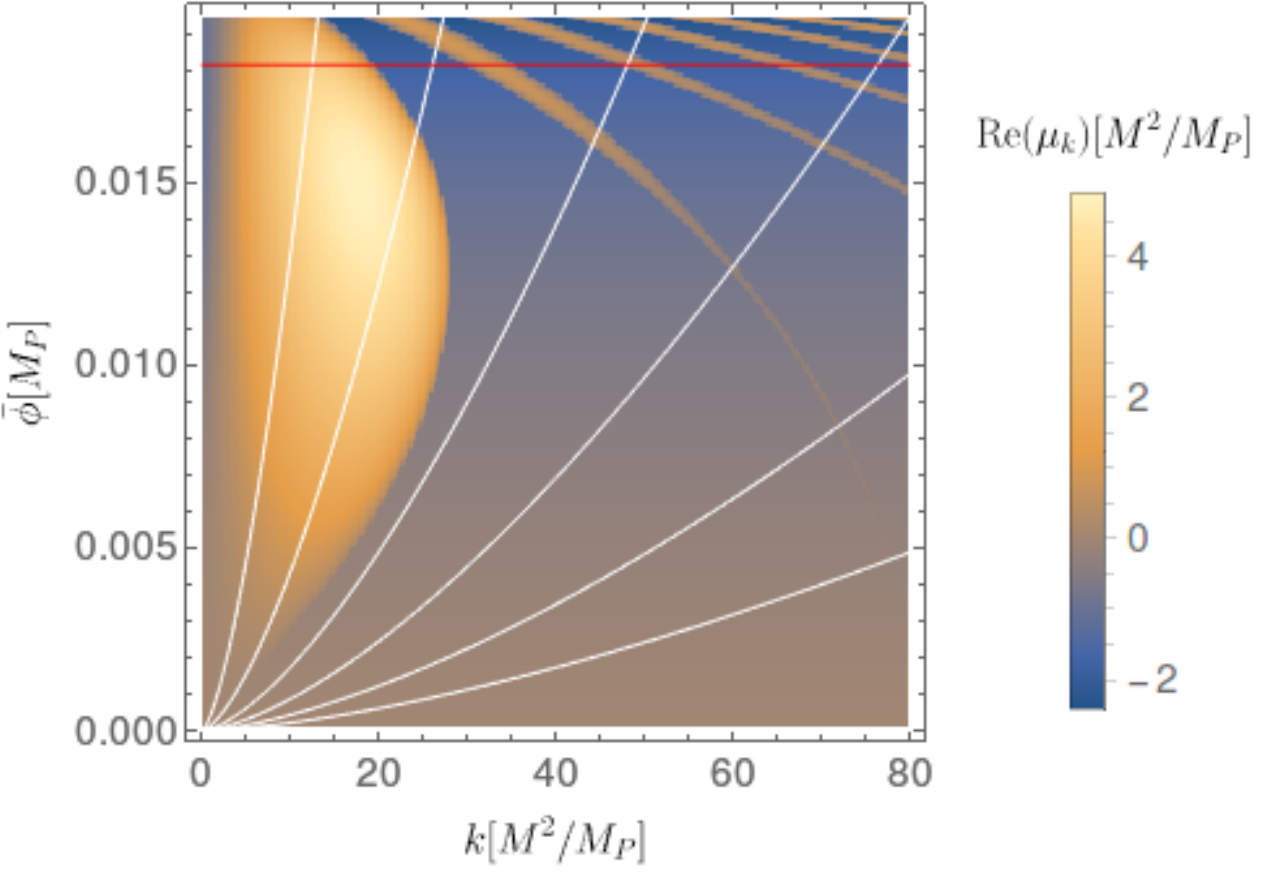}
\end{tabular}
\caption{\it Floquet exponents for the model WRI
for two different parameter choices consistent with Planck data.
  \label{fig:wri}}
\end{figure}

The models discussed in this Section are consistent with Planck data at $2\sigma$ confidence level and, since the potential vanishes at the minimum, we can perform Floquet analysis.
We did it for appropriately chosen parameters and we list our results for one- and two-parameter models in Tables~\ref{tab:ok1} and~\ref{tab:ok2}, respectively. 

\begin{table}
\begin{center}
{\footnotesize
\begin{tabular}{|l|l|l|r|}
\hline
name and acronym & potential & values of parameters & \parbox{1.5cm}{maximal Floquet exponent} \\
\hline
\parbox{3.4cm}{Generalized renormalizable point inflation \\ (GRIPI)} & \parbox{5cm}{$V(\phi)=M^4\bigg(\left(\frac{\phi}{\phi_0}\right)^2-\frac{4\alpha}{3}\left(\frac{\phi}{\phi_0}\right)^3+  \phantom{AAAAA}+\frac{\alpha}{2}\left(\frac{\phi}{\phi_0}\right)^{4}\bigg)$} &
 \parbox{4cm}{$1-\alpha=10^{-11},\;\phi_0=0.05M_{P}$ \\ $1-\alpha=10^{-11},\;\phi_0=0.07M_{P}$ } &  \parbox{1.5cm}{7.5 \\ 5.2 } 
 \\
\hline
\parbox{3.4cm}{KKLT inflation \\ (KKLTI)} & $V(\phi)=M^4\bigg(1+\Big(\frac{|\phi|}{\mu}\Big)^{-p}\bigg)^{-1}$ &
 \parbox{4cm}{ $\mu=0.01M_P,\; p=2$ \\ $\mu=0.01M_P,\; p=3$ \\ $\mu=0.01M_P,\; p=4$ \\ $\mu=0.04M_P,\; p=3$ } &  \parbox{1.5cm}{32 \\ 39 \\ 48 \\ 7.6 } 
 \\
\hline
\parbox{3.4cm}{T-models of \mbox{$\alpha$-attractor} inflation ($\alpha$TI)} & $V(\phi)=M^4\,\mathrm{tanh}^{2n}\!\left(\frac{|\phi|}{\sqrt{6\alpha}M_P}\right)$ &
 \parbox{4cm}{ $\alpha=10^{-4},\; n=1$ \\ $\alpha=10^{-4},\; n=2$ } & \parbox{1.5cm}{14 \\ 13 } 
 \\
\hline
\end{tabular}
}
\caption{\it Models with two parameters and large positive Floquet exponents. Values of Floquet exponents are given in units of $M^2/M_P^2$. \label{tab:ok2}}
\end{center}
\end{table}

GRIPI is in agreement with Planck data at $2\sigma$ confidence level for $\phi_0/M_{P}\leq 1$ and $0<1-\alpha<10^{-9}$. The fine-tuning of parameter $\alpha$ is necessary to obtain sufficient amount of inflation, but it is significantly less severe than in case of GMSSMI, so we decided to not to discard GRIPI from Floquet analysis.
We would also like to mention that the KKLTI with $p=2$ is equivalent to RGI;
in particular, our chosen parameter $\mu=0.01M_P$ corresponds to $\alpha=0.0001$, so we in fact consider two different examples. In the following, we
will formally distinguish between these two models, but one has to remember that this distinction is artificial.

As already mentioned, 
even in inflationary models exhibiting a strong instability of the perturbations there is no guarantee of efficient reheating, especially if the inflationary potential
is approximately quadratic near the minimum and has a plateau or flattens away from the minimum. In this case, long-lived oscillons are created \cite{AEFFH,AL0,AL,AS} and the equation of state
of the Universe is that of pressureless dust. Our short investigation of the models with positive Floquet exponents suggests that the creation and domination of the long-lived oscillons can be expected for the following models: GRIPI, MHI, RGI, WRI, KKLTI with $p=2$ and $\alpha$TI with $n=1$. Thus our analysis points towards KKLTI with $p\neq 2$ as the only new candidate for an inflationary model 
with self-resonance responsible for reheating, in addition to the already known example of $\alpha$TI with $n\neq1$ \cite{AL0,AL}.

\begin{figure}
\centering
\begin{tabular}{cc}
$\phi_0 = 0.07,\;1-\alpha=10^{-11}$
&
$\phi_0 = 0.05,\;1-\alpha=10^{-11}$
\\
\includegraphics[width=0.48\textwidth, height=0.35\textwidth]{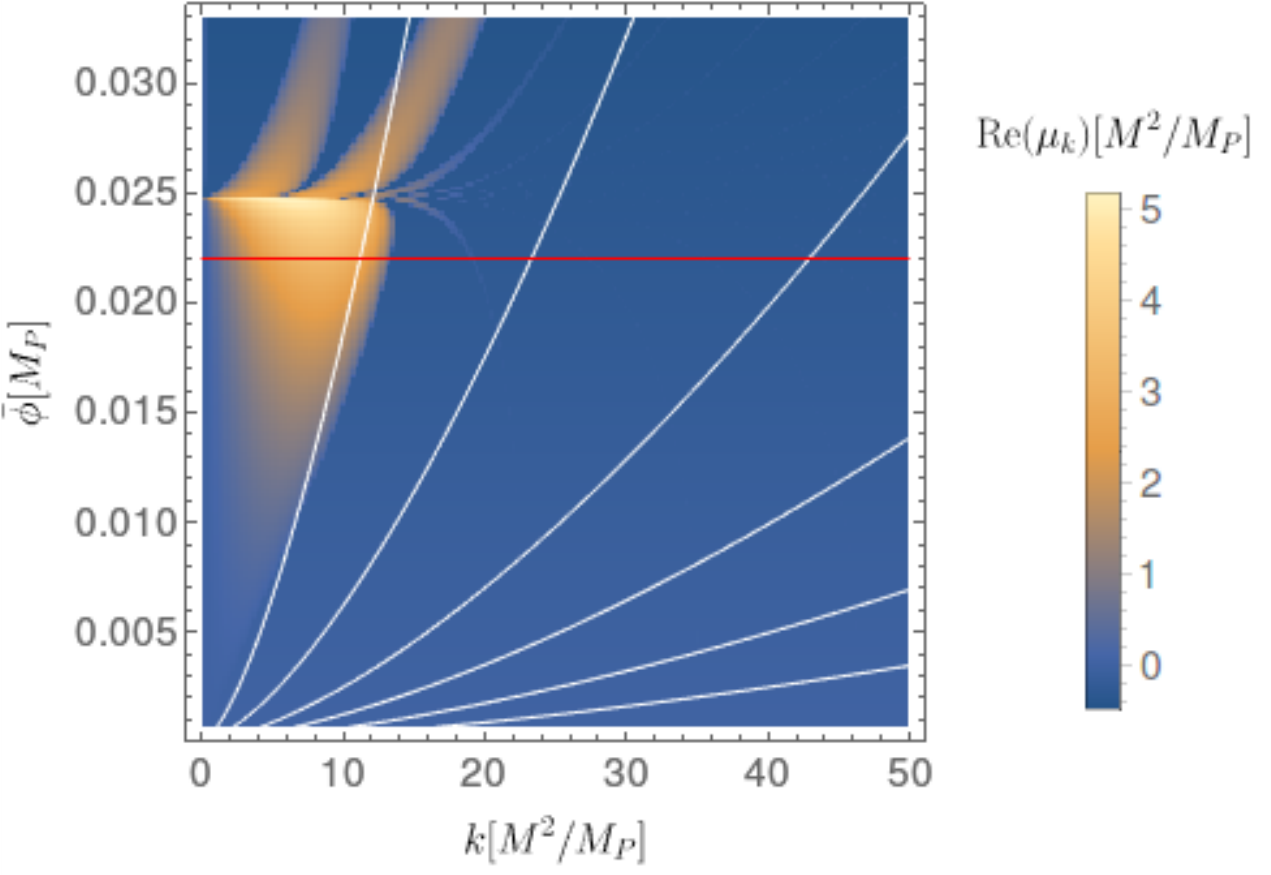}
&
\includegraphics[width=0.48\textwidth, height=0.35\textwidth]{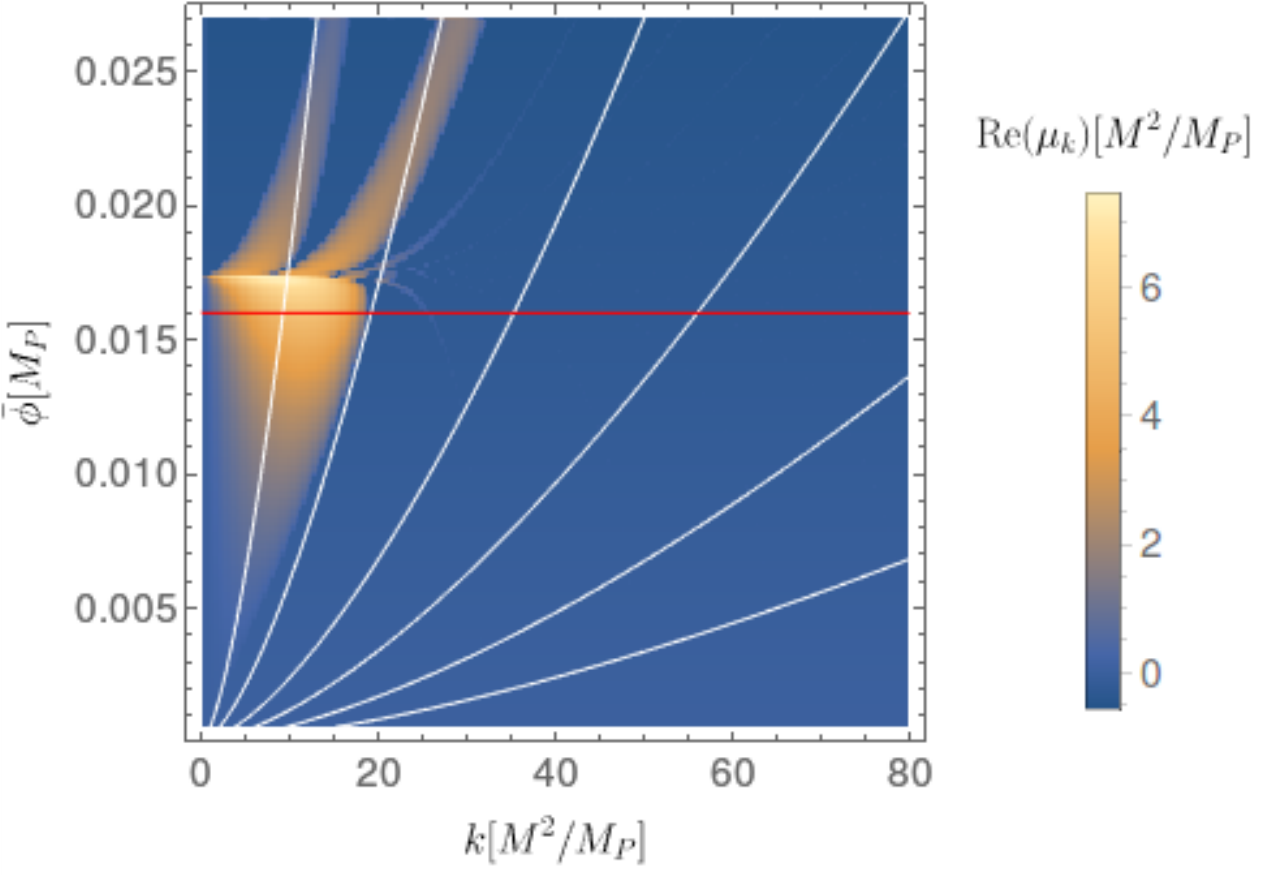}
\end{tabular}
\caption{\it Floquet exponents for the model GRIPI
for two different parameter choices consistent with Planck data.
  \label{fig:gripi}}
\end{figure}

\begin{figure}
\centering
\begin{tabular}{cc}
$\mu = 0.04,\; p=3$ & $\mu = 0.01,\; p=2$ \\
\includegraphics[width=0.48\textwidth, height=0.35\textwidth]{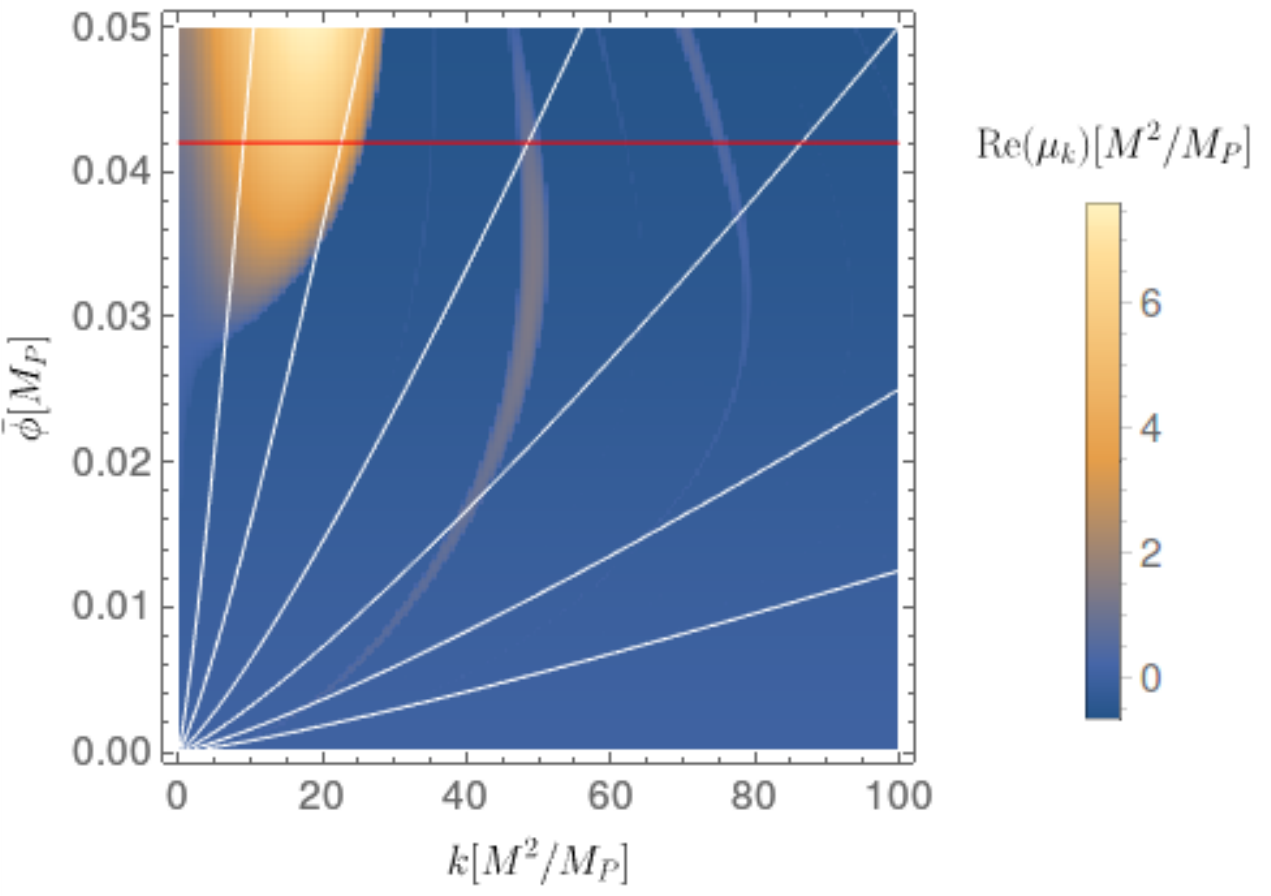} &
\includegraphics[width=.48\textwidth, height=0.35\textwidth]{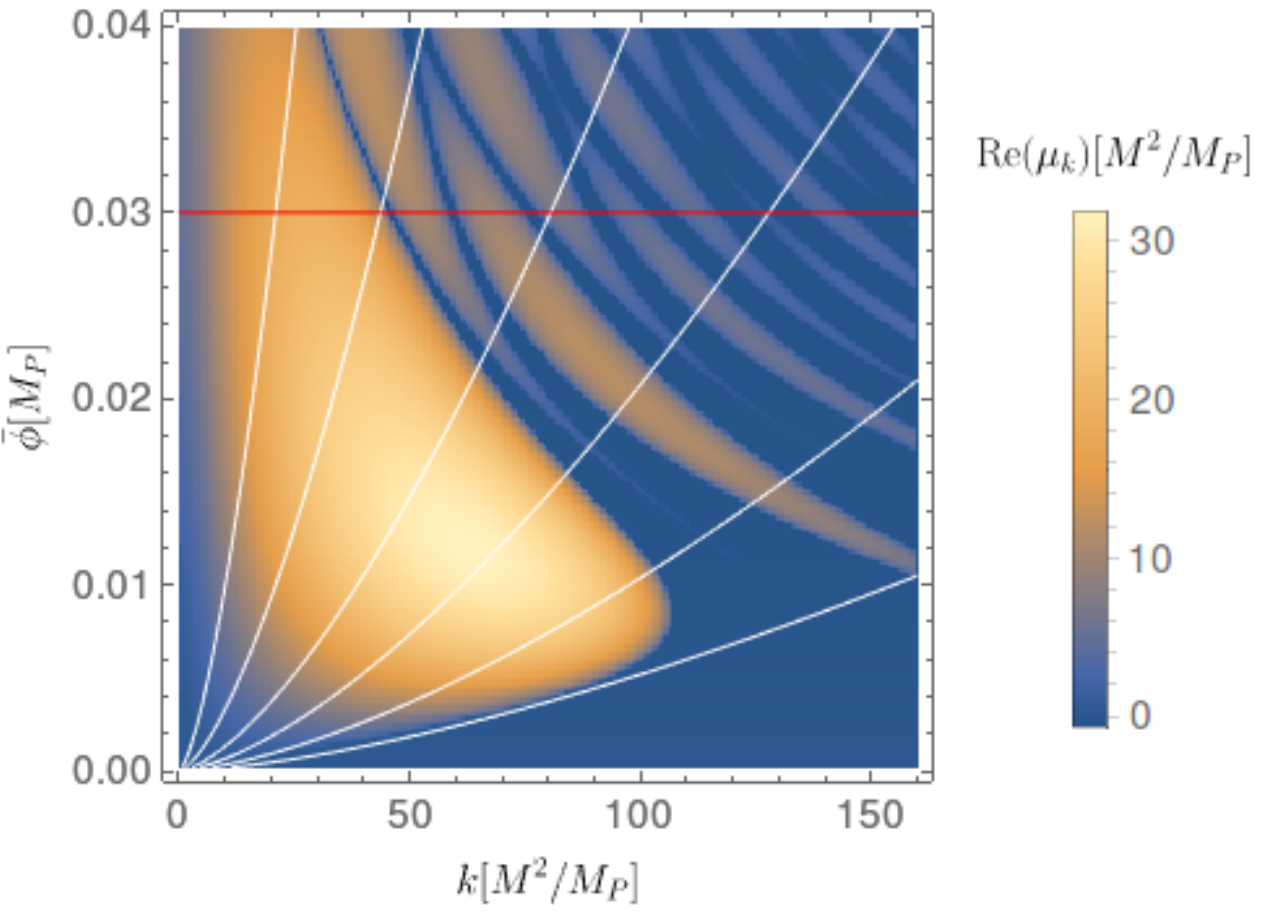} \\
& \\
$\mu = 0.01,\; p=3$ & $\mu = 0.01,\; p=4$ \\
\includegraphics[width=0.48\textwidth, height=0.35\textwidth]{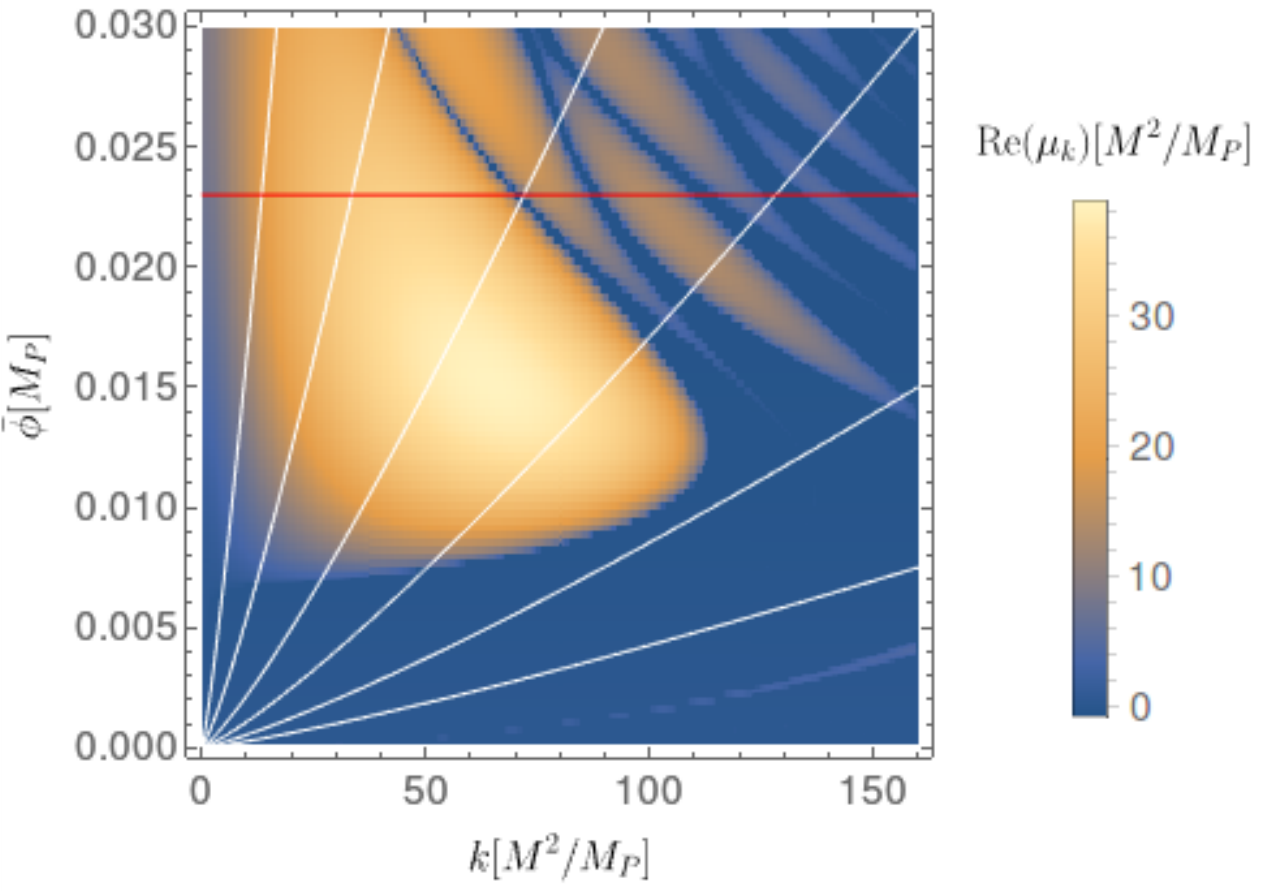} &
\includegraphics[width=0.48\textwidth, height=0.35\textwidth]{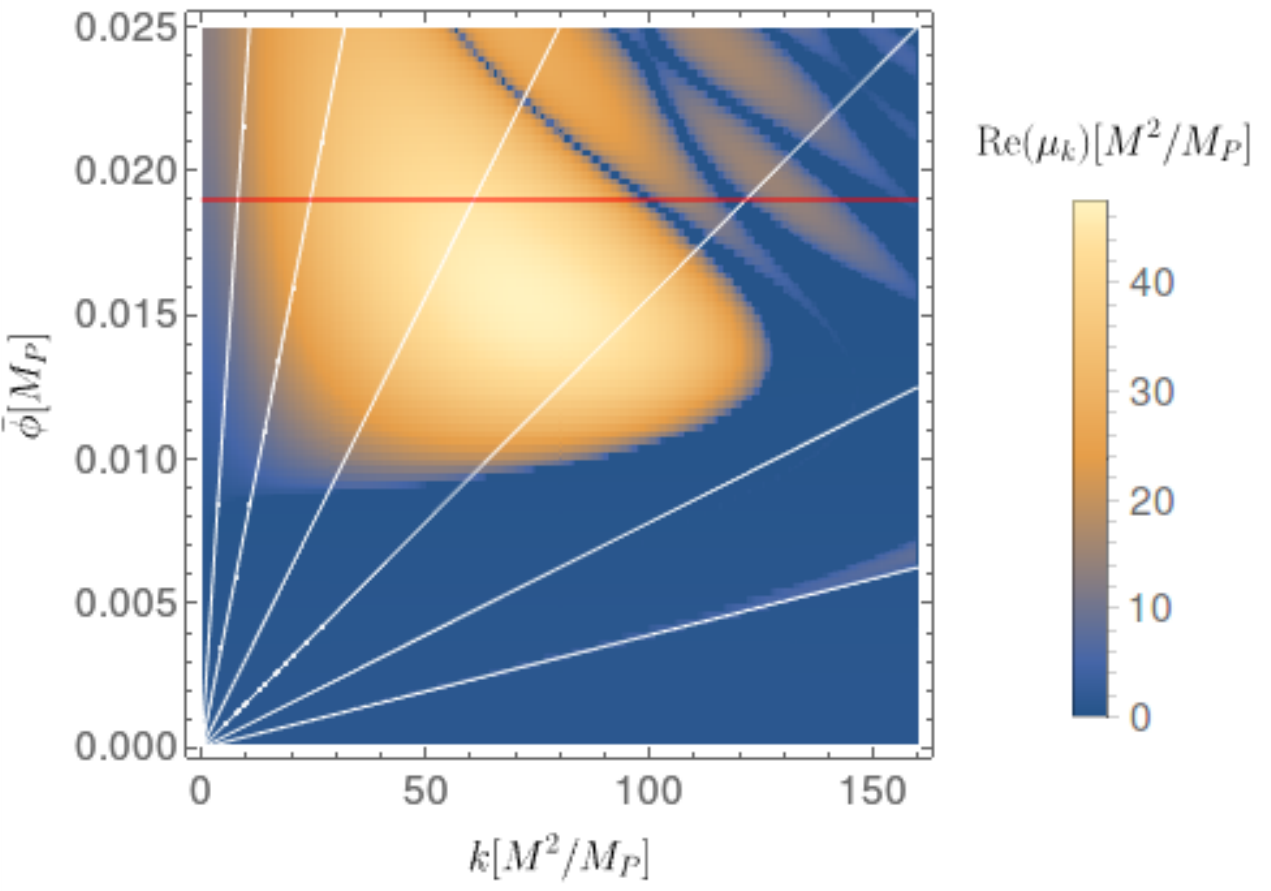}
\end{tabular}
\caption{\it Floquet exponents for the model KKLTI
for four different parameter choices consistent with Planck data.
  \label{fig:kklti}}
\end{figure}

\begin{figure}
\centering
\begin{tabular}{cc}
$n = 1,\; \alpha=10^{-4}$ & $n=2,\; \alpha=10^{-4}$ \\
\includegraphics[width=0.48\textwidth, height=0.35\textwidth]{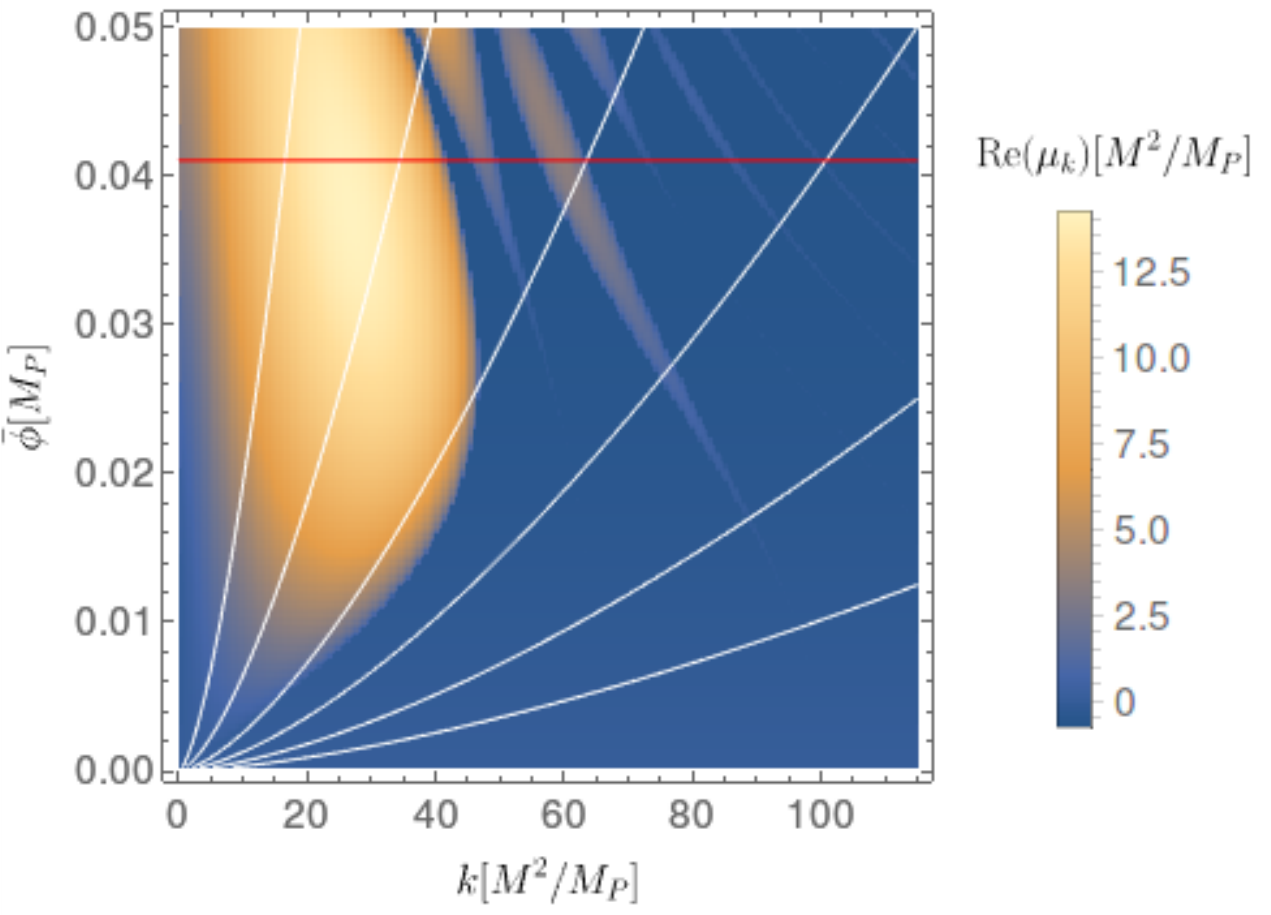} &
\includegraphics[width=.48\textwidth, height=0.35\textwidth]{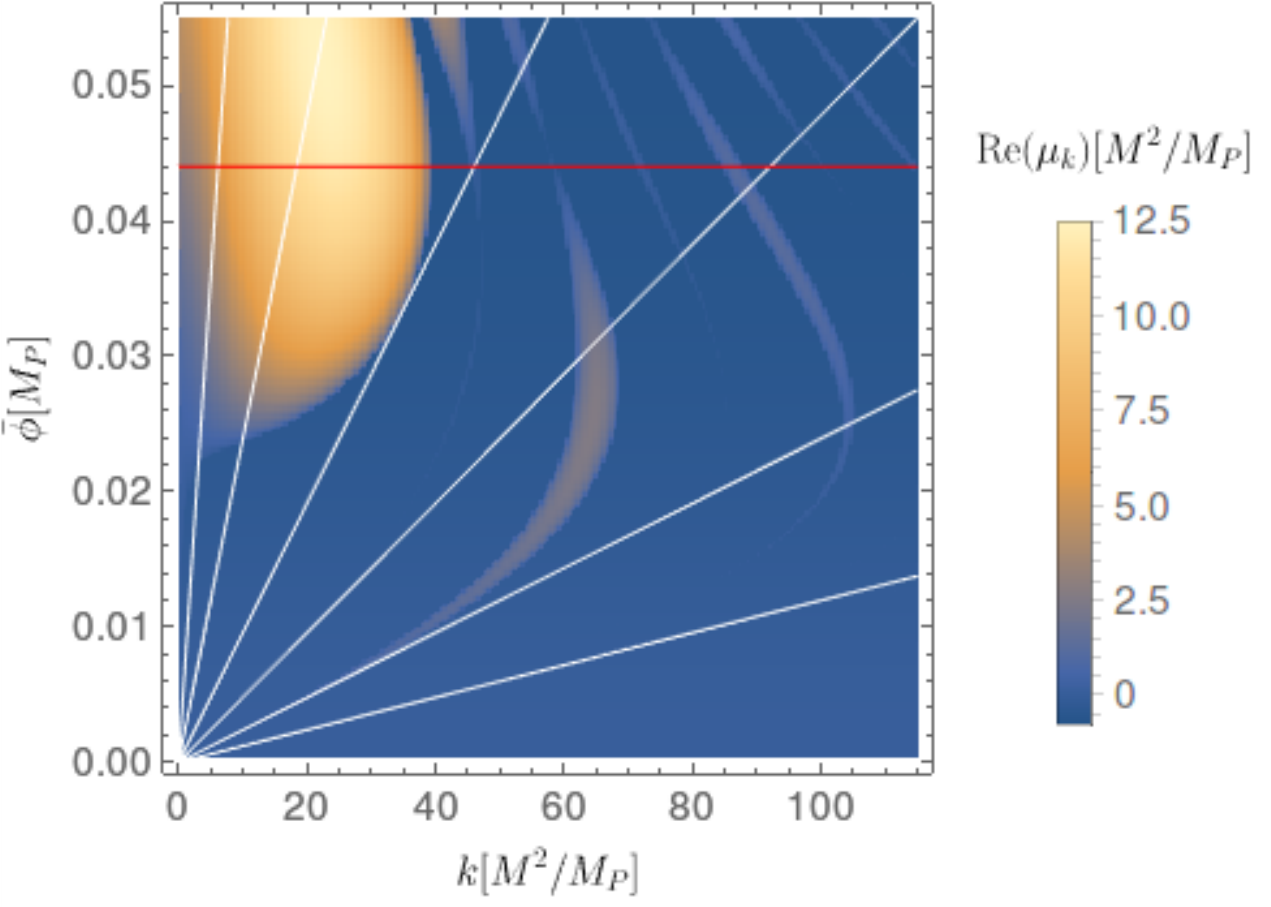} 
\end{tabular}
\caption{\it Floquet exponents for the model $\alpha$TI
for two different parameter choices consistent with Planck data.
  \label{fig:ati}}
\end{figure}

\begin{figure}
\centering
\begin{tabular}{cc}
$p=2,\;\mu=0.01M_P$
&
$p=3,\;\mu=0.01M_P$
\\
\includegraphics[width=0.48\textwidth, height=0.3\textwidth]{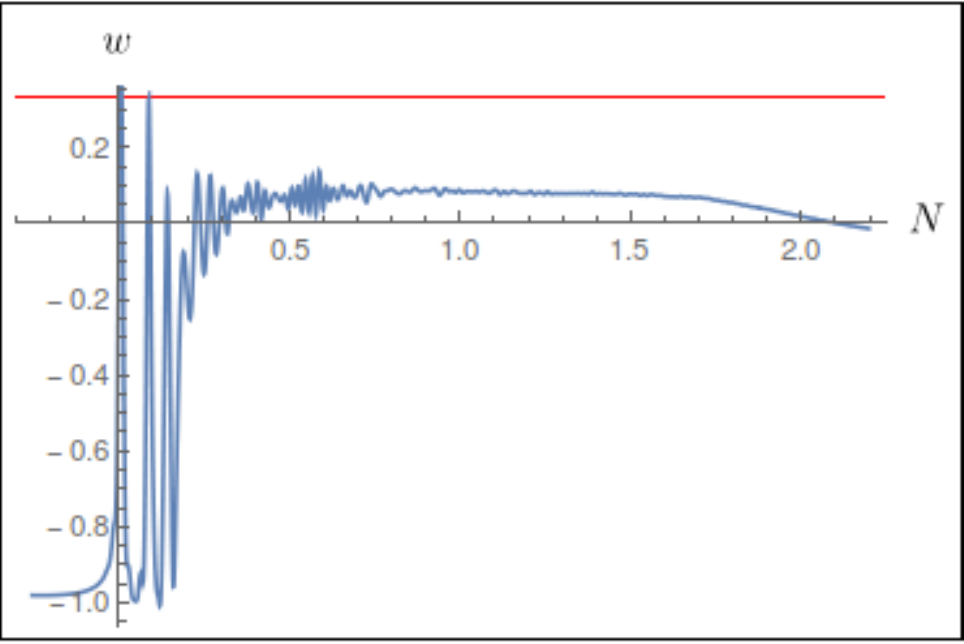}
&
\includegraphics[width=0.48\textwidth, height=0.3\textwidth]{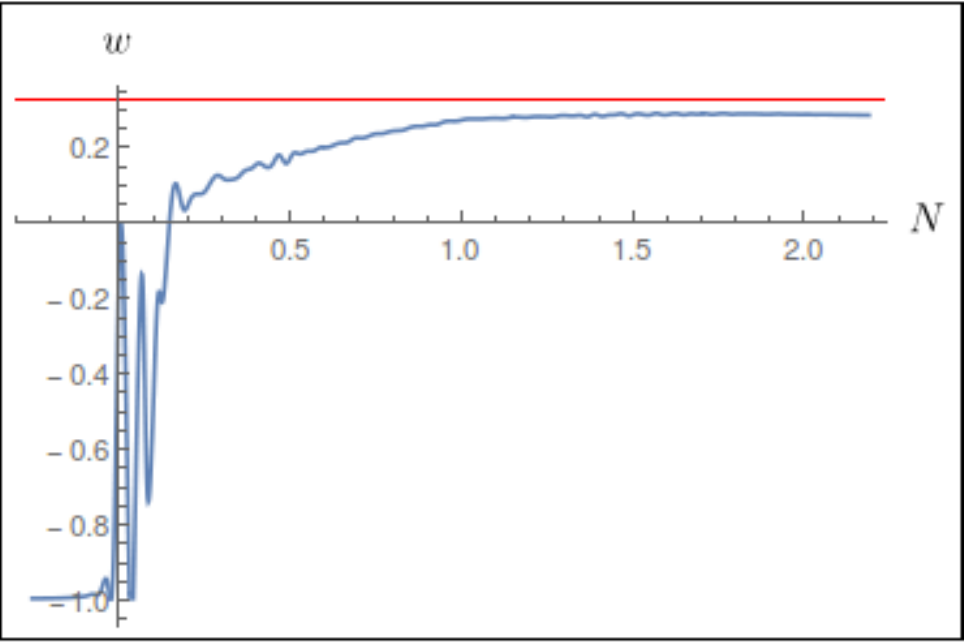}
\end{tabular}
\caption{\it Barotropic parameter $w=p/\rho$ for two realizations of KKLTI as a function of the number of efolds after the end of inflation. 
Horizontal red lines correspond to the equation of state characteristic of radiation.
  \label{fig:wplot}}
\end{figure}

\begin{figure}
\centering
\begin{tabular}{cc}
$p=2,\;\mu=0.01M_P$
&
$p=3,\;\mu=0.01M_P$
\\
\includegraphics[width=0.48\textwidth, height=0.3\textwidth]{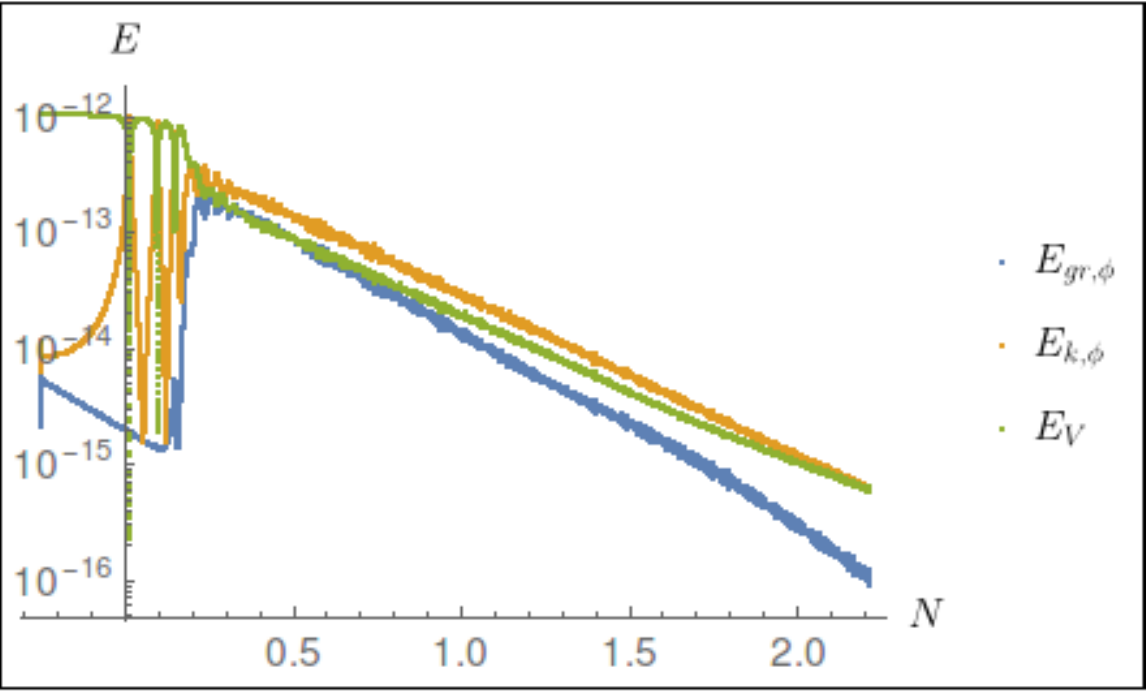}
&
\includegraphics[width=0.48\textwidth, height=0.3\textwidth]{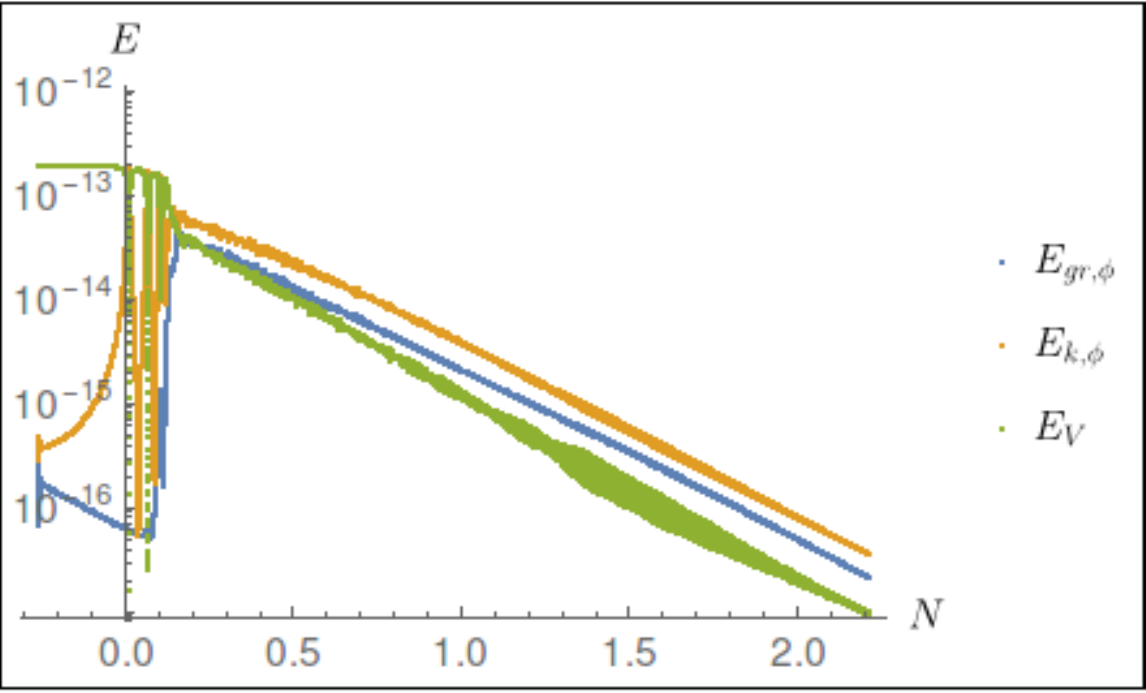}
\end{tabular}
\caption{\it Kinetic, gradient and  potential energy densities for two realizations of KKLTI as a function of the number of efolds after the end of inflation. 
  \label{fig:budget}}
\end{figure}

In order to corroborate this conclusion, we
performed numerical lattice simulations 
of the evolution of the perturbations of the inflaton field in KKLTI with $p=2$ and $p=3$,
using a code described in \cite{KTW}.
We used cubic lattices of a linear size $N_\mathrm{lattice}=64$ with the momentum space cutoff $k_\mathrm{max}=1500M^2/M_P$;
we expect that the latter is a good trade-off between granularity in Floquet instability regions and the necessity of including high frequency modes.
Our results for the evolution of the barotropic parameter $w=p/\rho$ are shown in Figure \ref{fig:wplot} and the results for the evolution 
of various time-averaged components of the energy budget of the system are shown in Figure \ref{fig:budget}. We would like to stress that after a fraction of an 
efold the evolution of the inflaton field enters a nonlinear regime and the numerical results cannot be directly compared to the Floquet analysis.

For $p=2$, the gradient energy, which corresponds to inhomogeneities, is initially created, but it decays more quickly than the potential and kinetic energy.
As a result, after just a couple of efolds, the inhomogeneities are a subdominant component of the total energy and the barotropic parameter approaches zero.
On the contrary, for $p=3$ the gradient energy is significantly larger than potential energy, showing that the excited modes of the inflaton dominate the post-inflationary
Universe. Less than a third of the energy density remains in the form of potential density after two efolds and the resulting effective barotropic parameter approaches 0.3, interpolating between
the value of 1/3, characteristic of radiation, and 1/5, characteristic of a condensate oscillating in potential $\sim|\phi|^3$, proportionately to the contributions of the two energy component to the total energy budget.
The potential energy density does not decay away completely, which suggests that a there remains a fraction of the homogeneous inflaton condensate. 
However, a deviation by 10\%
of the barotropic parameter $w$ from 1/3 translates only to 1\% shift of the inflationary observables, the scalar spectral index and the tensor-to-scalar ratio, compared to the pure radiation
with $w=1/3$ \cite{AL}. In this sense, creation of inhomogeneities of the inflaton field at the end of inflation effectively plays the role of reheating in this example and greatly reduces uncertainties related to the details 
of the reheating era.

\subsection{Floquet analysis and the expansion of the Universe}

\begin{figure}
\centering
\begin{tabular}{cc}
\includegraphics[width=0.48\textwidth, height=0.25\textwidth]{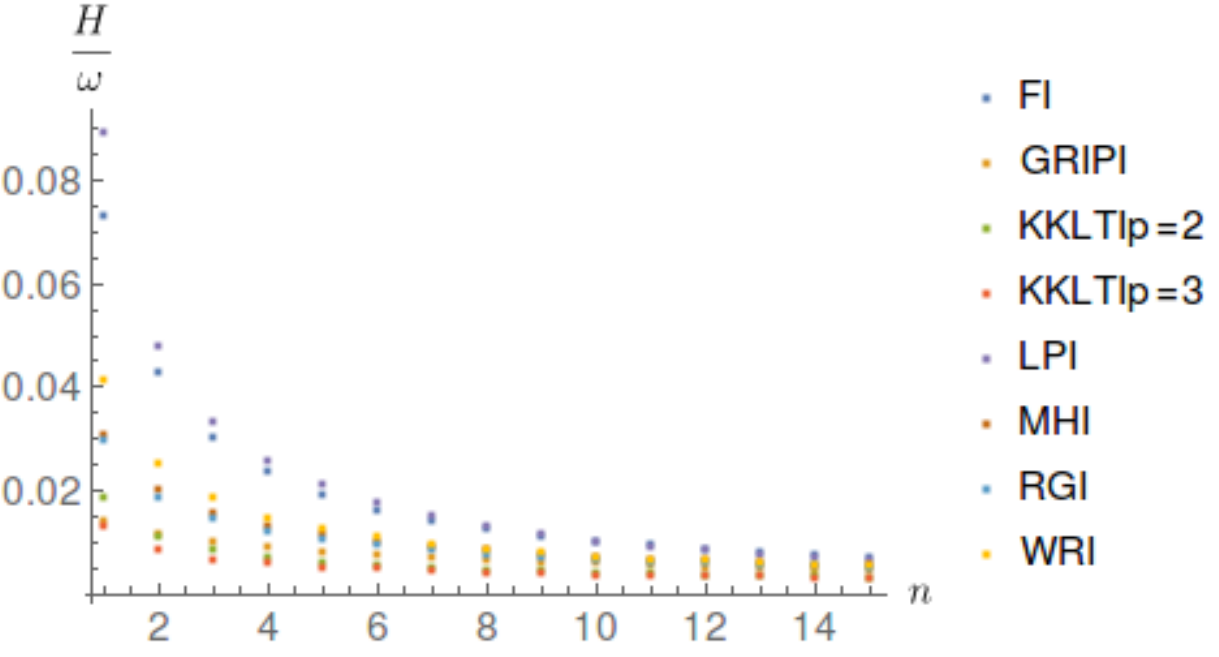} &
\includegraphics[width=.48\textwidth, height=0.25\textwidth]{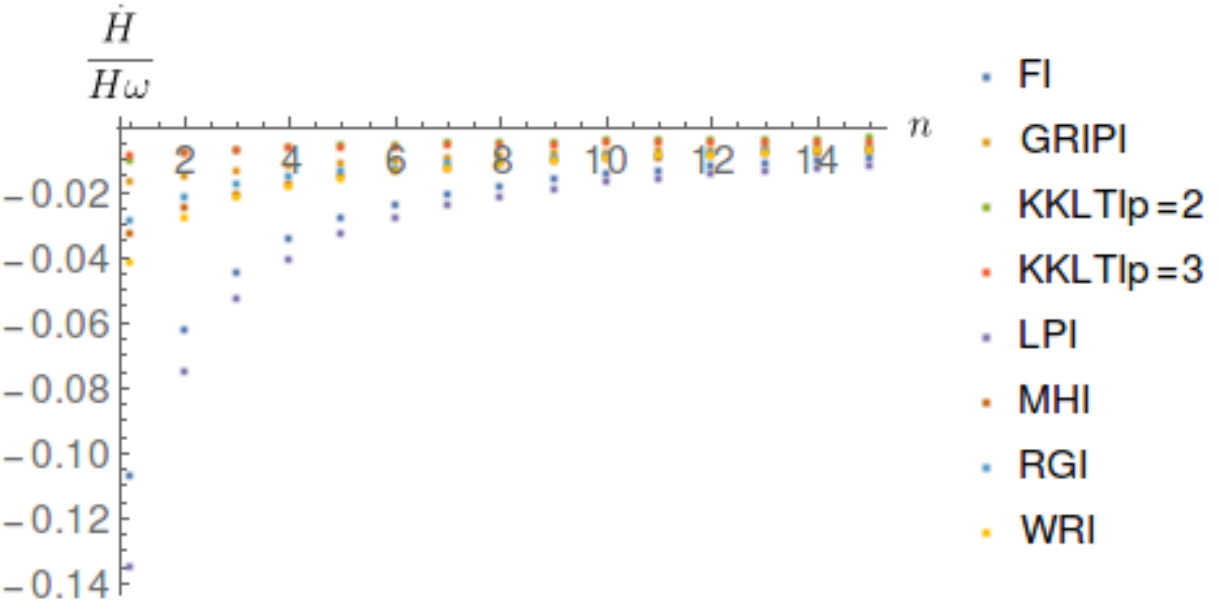} 
\end{tabular}
\caption{\it Values of $H/\omega$ and $\dot{H}/H\omega$ for the first 15 oscillations of the inflaton field around the minimum of the potential for selected models considered in Sections \ref{sec:small} and~\ref{sec:large}.
  \label{fig:ex1}}
\end{figure}

The formalism of the Floquet theorem applied to cosmological setting requires that the expansion of the Universe, quantified by the scale factor $a(t)$ and the Hubble parameter $H(t)$,
is slow compared to a typical frequency of the oscillations of the inflaton field around the minimum of the potential. To this end, we verified that the logarithmic time derivatives of these quanities,
$H$ and $\dot{H}/H$, respectively, are in fact significantly smaller that the frequency scale of the oscillations, defined by $\omega=2\pi/T$, where $T$ is the duration of a given oscillation.
We present our results in Figure \ref{fig:ex1} for several models of inflations discussed in this Section and we find that this is indeed the case: 
$H/\omega$ and $\dot{H}/H\omega$ do not normally exceed a few per cent for the first 15~oscillations. Incidentally, these results also show that the expression for the mass of the perturbations (\ref{muphi2}) is dominated
by the first contributions; other contributions originate from metric perturbations and are suppressed by $M_P$. As our lattice simulations do not include metric perturbations, this means that we do not introduce
significant discrepancies by using the full result (\ref{muphi2}) for our Floquet analysis.

\section{Summary}
\label{sec:five}
We performed 
Floquet
analysis of the evolution of the perturbations of the inflaton
after the end of inflation
in a number of  theoretically justified and observationally favored single-field inflationary models considered by other authors.
We showed that these perturbations can be unstable and lead to to the fragmentation of inflaton condensate in six of investigated models. However, we found  that, in addition to the already known example of $\alpha$TI with $n\neq1$, only in case of KKLT inflation with $p\neq 2$, self-resonance can be a good scenario for reheating, {\it i.e.} describe the transition of the equation of state of the Universe close to that of radiation. 
In the remaining cases, the instability leads to creation of long-lived oscillons, for which the typical equation of state is that of non-relativistic matter. Therefore, in those cases, other scenario of reheating are necessary.
Taken together,
our findings suggest that efficient reheating {\it via} self-resonance, although possible in principle, is rather rare among inflationary models.

\subsubsection*{Acknowledgements}

The authors are supported by grant No.\ 2014/14/E/ST9/00152 from the National Science Centre (Poland).

\end{document}